# Reorientable Spin Direction for Spin Current Produced by the Anomalous Hall Effect

Jonathan D. Gibbons,[1] David MacNeill,[1] Robert A. Buhrman,[1] and Daniel C. Ralph[1,2]

[1]Cornell University, Ithaca, NY 14853, USA

[2]Kavli Institute at Cornell for Nanoscale Science, Ithaca, New York 14853, USA

We show experimentally that the spin direction of the spin current generated by spin-orbit interactions within a ferromagnetic layer can be reoriented by turning the magnetization direction of this layer. We do this by measuring the field-like component of spin-orbit torque generated by an exchange-biased FeGd thin film and acting on a nearby CoFeB layer. The relative angle of the CoFeB and FeGd magnetic moments is varied by applying an external magnetic field. We find that the resulting torque is in good agreement with predictions that the spin current generated by the anomalous Hall effect from the FeGd layer depends on the FeGd magnetization direction $\hat{m}_{\text{FeGd}}$ according to $\vec{\sigma} \propto \left( \hat{y} \cdot \hat{m}_{\text{FeGd}} \right) \hat{m}_{\text{FeGd}}$, where $\hat{y}$ is the in-plane direction perpendicular to the applied charge current. Because of this angular dependence, the spin-orbit torque arising from the anomalous Hall effect can be non-zero in a sample geometry for which the spin Hall torque generated by non-magnetic materials is identically zero.

Spin transfer torques exerted by spin currents arising from spin-orbit interactions have the potential to provide greatly improved efficiency in the manipulation of nano-magnetic memory bits. Strong spin-orbit interactions in nonmagnetic heavy metals, for example, can give rise to the spin Hall effect (SHE), which causes electrons with opposite spins to be deflected in opposite transverse directions [1-3]. As a result, when charge current is applied in the plane of a heavy metal/ferromagnet bilayer the SHE drives injection of spins from the heavy metal into the ferromagnet, generating a spin-transfer torque acting on the magnetization direction, $\hat{m}_{\text{sensor}}$ [4,5]. In typical heavy metal/ferromagnet bilayers the reflection and rotational symmetries of the sample require that the net orientation $\vec{\sigma}$ of the injected spins is transverse to both the charge current flow and the interface normal. As a consequence, the spin transfer torque generated by the SHE in such samples is restricted to consist just of an anti-damping component that points strictly in-plane (of the form $\propto \pm \hat{m}_{\text{sensor}} \times (\vec{\sigma} \times \hat{m}_{\text{sensor}}))$ plus a field-like torque (of the form $\propto \pm \vec{\sigma} \times \hat{m}_{\text{sensor}}$) [6,7]. This restriction can be detrimental for applications; an anti-damping torque that lies in the sample plane is incapable of driving highly-efficient anti-damping switching of devices with perpendicular magnetic anisotropy [8]. The symmetry requirement that mandates the fixed direction of $\vec{\sigma}$ can be relaxed by using single-crystal spin-orbit materials with sufficiently low crystal symmetries [7], but it will be difficult to incorporate such materials into practical technologies.

Electrons inside ferromagnetic metals, like those in non-magnetic heavy metals, can also undergo spin dependent deflection. This deflection produces the well-known anomalous Hall effect (AHE) [9], and is also expected to create charge-current/spin-current interconversion in ferromagnets analogous to the SHE and inverse spin Hall effect (ISHE) in non-magnetic metals. In fact the ISHE has already been observed in a variety of ferromagnetic materials [10-15]. However, the transverse spin currents arising from spin-orbit interactions within a ferromagnet are predicted to have a qualitatively different character than the SHE in a nonmagnetic heavy metal due to the presence of the strong ferromagnetic exchange field [16,17]. Spins in a ferromagnet precess rapidly around the magnetization (exchange-field) direction, so that any net macroscopic spin current within a ferromagnetic layer should have the spin polarized along $\pm\hat{m}_{source}$, where $\hat{m}_{source}$ is the magnetic-moment-orientation of the source layer. This suggests that it should be possible to reorient the polarization of the spin current produced by spin-orbit interactions within a ferromagnet by reorienting $\hat{m}_{source}$, to thereby gain the ability to reorient at will both the anti-damping torque and the field-like torque that the spin current applies to a second magnetic layer. Here, we report measurements of a spin-orbit-induced spin current generated from a source magnetic layer, detected by measuring the field-like spin-transfer torque applied to a second, spin-absorbing sensor magnetic layer (Fig. 1a). We observe the predicted [16,17] reorientation of the injected spins as the source-layer moment is rotated in the sample plane. As one consequence, we show that the spin current generated by a ferromagnetic source layer is able to apply spin-transfer torque in a sample configuration where the conventional spin Hall torque produced by a non-magnetic heavy metal is zero.

For our experiments, we use a thin-film stack comprising a 10 nm IrMn layer, followed by a 4 nm $Fe_{95}Gd_5$ (henceforth FeGd) source layer, a 2 nm Hf spacer, and a 2 nm $Co_{40}Fe_{40}B_{20}$ sensor layer (henceforth CoFeB) capped with 3 nm of Hf. Our films are grown on sapphire wafers via DC magnetron sputtering, annealed at 420 K for one hour in a 0.2 T in-plane magnetic field to set the exchange bias direction of the IrMn layer, and then patterned into 120 μm by 20 μm Hall bars with 5 μm voltage probes using optical lithography and ion milling, with the current direction aligned with the exchange bias direction (Fig. 1b). We choose FeGd as our spin-source material because rare earth ferromagnetic alloys have the potential for efficient spin-current generation – in particular, past research[18] has found that certain iron-gadolinium alloys may exhibit a strong anomalous Hall effect. (We have not yet attempted optimization of the Gd concentration.) The exchange bias from IrMn acting on the FeGd layer allows us to control the angle between the magnetic moments of the CoFeB and FeGd layers, and therefore to study whether the orientation $\vec{\sigma}$ of the spin current produced by current flow in the FeGd layer depends on the FeGd moment orientation. To obtain the most accurate control over this offset angle, the samples are designed to have in-plane magnetic anisotropy and our external magnetic fields are also applied in-plane; the soft CoFeB layer saturates along even weak external fields, whereas the FeGd layer rotates smoothly from the exchange bias direction to the applied field direction as the strength of the external field is increased. The exchange bias grows with decreasing temperatures, so we performed all measurements at cryogenic temperatures, approximately 30 K. The 30 K resistivities of the various layers, determined by measurements of separate test samples, are approximately 209 ± 20 μΩcm (IrMn), 64 ± 8 μΩcm (FeGd), 94 ± 35 μΩcm (CoFeB).

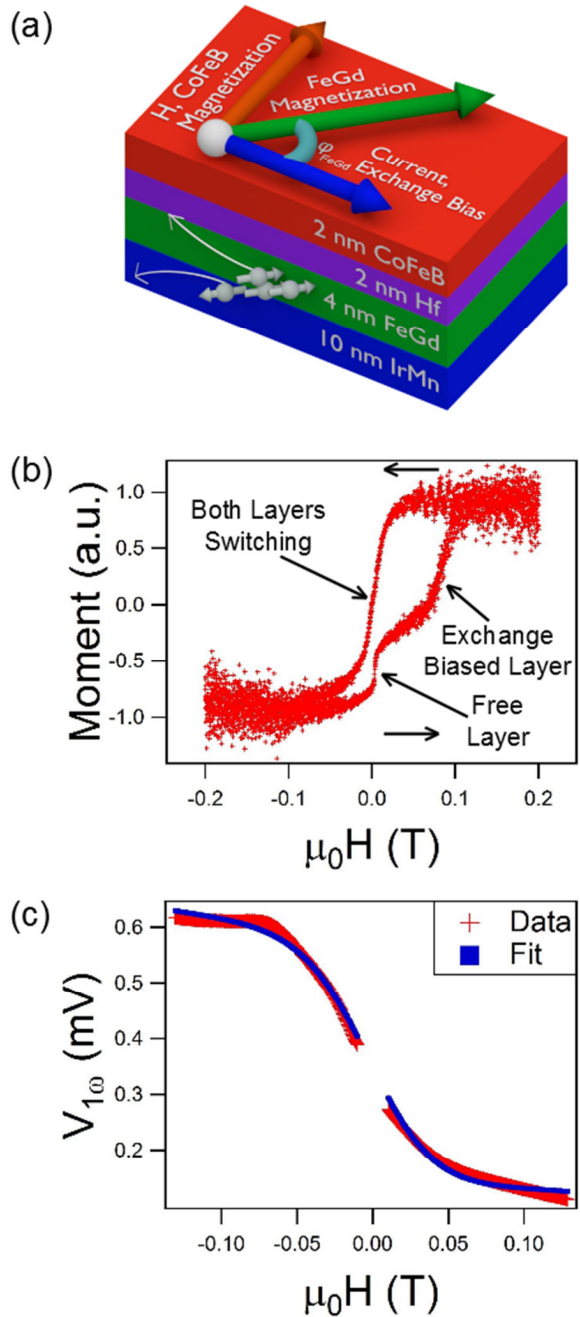

Fig. 1. (a) Schematic of the device geometry. (b) Magnetization of an unpatterned IrMn/Fe$_{95}$Gd$_5$/Hf/Co$_{40}$Fe$_{40}$B$_{20}$/Hf multilayer at 30 K measured using vibrating sample magnetometry, showing the exchange-biased switching of the FeGd layer with high coercivity, and the low-coercivity switching of the CoFeB layer. (c) First harmonic Hall data taken at 30 K, with the exchange bias parallel to the current and the magnetic field perpendicular to the exchange bias. The value of the FeGd layer exchange bias is extracted from the fit to the planar Hall signal.

Figure 1b shows a measurement of the magnetization of our unpatterned film stack as a function of a magnetic field applied in the sample plane parallel to the set exchange bias, as characterized by vibrating sample magnetometry (VSM) at 30 K. When increasing the magnetic field from zero, we see first the in-plane magnetization switching of the low-coercivity CoFeB sensor layer. This is followed at higher fields by the more gradual switching of the strongly exchange biased FeGd layer. These data verify that both magnetic layers have in-plane magnetic anisotropy.

To measure current-induced torques on the sensor layer (which can arise from either spin currents or an Oersted field) we use the second-harmonic Hall technique [6,19-22] in which a low-frequency (1000 Hz) alternating current is applied to the device and the induced Hall voltage is measured at the second harmonic frequency. For all samples, we apply a 5 V signal, so that the current density within a given material layer is approximately the same between samples. In principle, for an in-plane-magnetized sample the second-harmonic Hall technique can provide measurements of both the in-plane and out-of-plane components of current-induced torque, and can also distinguish anti-damping torques from field-like torques (see Table 1), but one must be careful to distinguish the spin-torque signals from artifacts associated with thermoelectric effects [21]. In-plane torques correspond to out-of-plane effective magnetic fields (by the right-hand rule) so that they tend to pull an in-plane sensor layer slightly out-of-plane, giving a second harmonic Hall voltage signal on account of mixing between an oscillating anomalous Hall resistance and the oscillating current. In this case, the in-plane torque competes with the magnetic anisotropy field $H_k$ of the thin-film CoFeB layer, and because the anisotropy field is much larger than the applied field ($H_k \gg H$) in our experiment, the magnitude of the second-harmonic Hall signal should be small and approximately independent of the magnitude of applied magnetic field for a fixed field orientation near $H = 0$. For the usual case of an anti-damping in-plane torque, flipping the magnetization direction of the sensor layer changes the sign of the deflection, so that the final second-harmonic Hall signal should have a sign change near $H = 0$ upon reversal of our low-coercivity CoFeB layer, and should otherwise be flat as a function of swept magnetic field (Table 1(a)). (In the presence of a field-like in-plane torque, reversing the sensor layer would not change the sign of the second-harmonic signal, Table 1(b).) Out-of-plane torques, on the other hand, correspond to in-plane effective fields, causing an in-plane rotation of the sensor layer's magnetization that is detected through mixing with the planar Hall resistance. This in-plane field competes only with the applied magnetic field. The size of the sensor layer deflection is then inversely proportional to the applied field, and we expect a second-harmonic Hall signal whose magnitude diverges as $1/|H|$. For a field-like out-of-plane torque, reversing the CoFeB magnetization direction changes the sign of the second harmonic signal, so that the signal should flip sign as the field is swept through zero to reorient the low-coercivity CoFeB layer (Table 1(c)), while for an anti-damping out-of-plane torque the signal would be proportional to $1/|H|$ with no sign change (Table 1(d)). All of the entries in Table 1 assume that the spin-current-induced torque is nonzero and approximately independent of magnetic field in the range near H=0; if the magnitude of the torque depends on field there will be deviations (as discussed below) from the ordinary $1/|H|$ dependence for out-of-plane torques and *H*-independent behavior for in-plane torques.

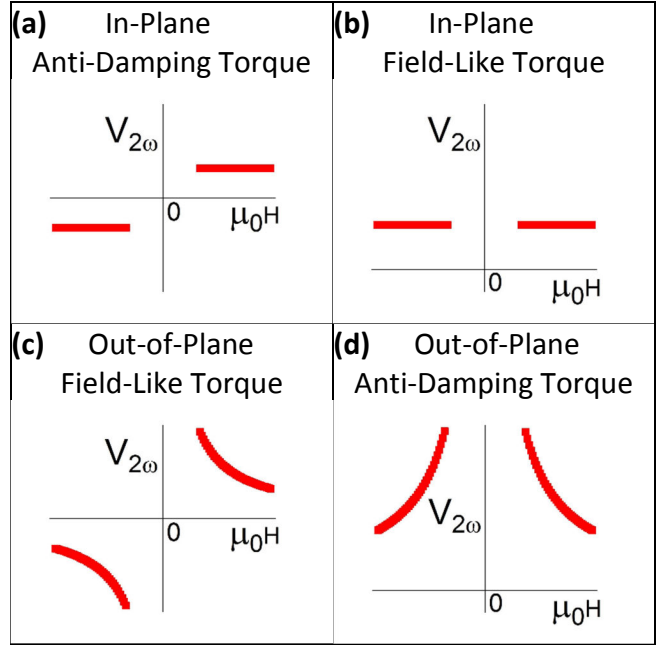

Table 1. Predicted second harmonic Hall signals resulting from various forms of spin torque acting on a ferromagnetic sensor layer with in-plane magnetization for applied fields near zero. These results assume that the spin torque is nonzero and does not vary strongly with applied magnetic field.

To best distinguish whether there is a spin-orbit torque arising from the magnetic FeGd layer that depends on the orientation of the FeGd moment, we consider a measurement configuration for which the torques arising from both the current-generated Oersted field and also any conventional spin Hall effect must be zero. We sweep the applied magnetic field perpendicular to the current flow direction (and therefore also perpendicular to the direction of the exchange bias), so that the low-coercivity CoFeB sensor layer is quickly saturated perpendicular to the current (for $|\mu_0 H|$ greater than approximately 0.01 Tesla), while the angle, $\varphi_{\text{FeGd}}$, of the FeGd moment rotates slowly away from the exchange bias direction with increasing field magnitude (see Fig. 1a). Because the sensor-layer moment is oriented transverse to the current flow direction (*i.e.*, along the Oersted field), there can be no Oersted torque. Likewise, in this geometry the sensor moment is also parallel to the spins that would be created by any conventional spin Hall effect, so that there can be no conventional spin Hall torque. This geometry has the further advantage that any possible artifacts from the anomalous Nernst effect or the longitudinal spin Seebeck effect in the CoFeB layer must also be zero (since for either mechanism $V_{Hall} \propto \nabla T \times \hat{m}$, and the thermal gradient, $\nabla T$, is assumed to be out-of-plane [21,23-25]). The variation of $\varphi_{FeGd}$ with applied field can be determined based on the first-harmonic Hall signal and calibration of the strength of the planar Hall effect, as shown in Fig. 1c. Using the equation: $\varphi_{FeGd} \approx \varphi_{FeGd}^{0} + \tan^{-1}(H/H_{ex})$ we determine that the FeGd magnetization angle rotates from $\varphi_{FeGd}$ = -58.6° through zero to 64.0° in this range of applied field. Fitting to the planar Hall first harmonic data allows us to extract the value of the exchange bias field as $\mu_0 H_{ex}$ = 0.071 ± 0.001 T.

Our experimental results in this geometry for the second harmonic Hall voltage as a function of the applied magnetic field (perpendicular to the exchange bias direction) are shown in Fig. 2a. We have excluded data for field magnitudes less than 0.01 Tesla from our analysis, because in this regime the CoFeB layer undergoes a spatially non-uniform reversal process that invalidates the macrospin analysis we use to interpret the second harmonic Hall measurements. We will exclude the same range of field for all data analyzed below from samples containing the CoFeB layer. We observe in Fig. 2a a substantial signal whose magnitude diverges approximately as $1/|H|$ as $H$ approaches zero, with a sign change as $H$ is swept through 0. This is the signature of an out-of-plane field-like torque (Table 1(c)). Because the low-coercivity CoFeB sensor is the only layer that reverses near $H$ = 0 (while the FeGd magnetization remains oriented near the exchange bias direction), this behavior indicates that the signal arises from a torque on the CoFeB sensor layer. The magnitude of this spin-current-induced torque is not constant, but rather changes as a function of changing field magnitude, and hence as a function of changing $\varphi_{FeGd}$. This is evident because if the magnitude of the torque were constant, the magnitude of the second-Harmonic hall signal should decrease monotonically with increasing field magnitude as $\propto 1/|H|$, while the data display a distinctly non-monotonic dependence at positive field, with $V_{2\omega}$ initially increasing and then decreasing as $\mu_0 H$ increases from 0.

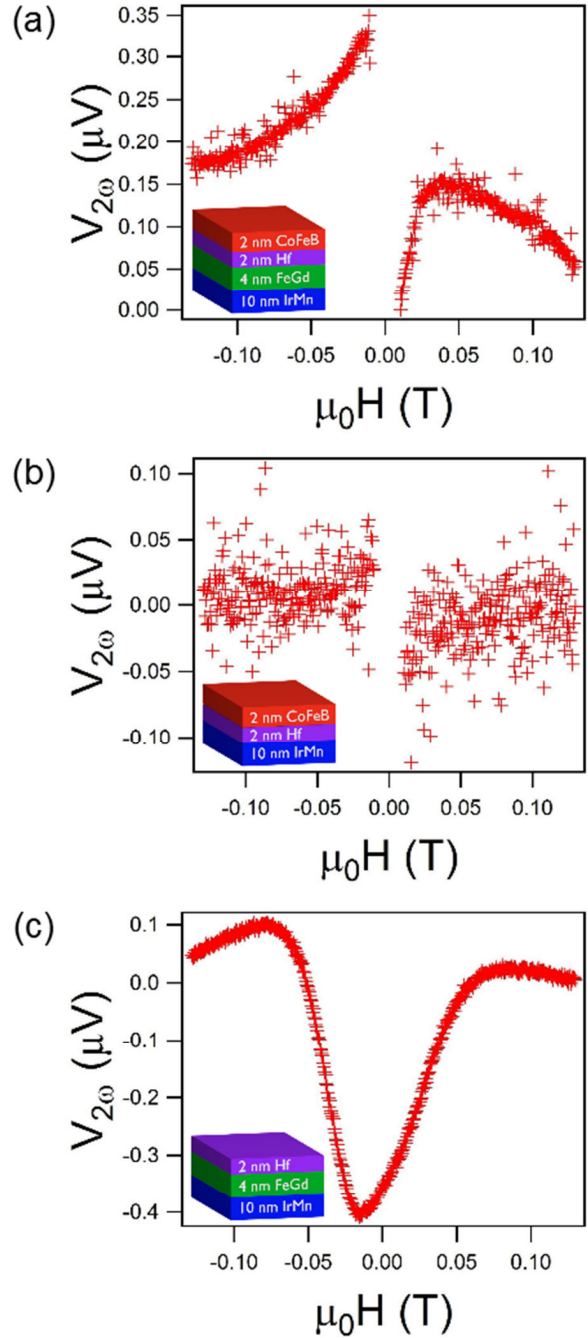

Fig. 2. (a) The measured second-harmonic Hall signal at 30 K for the full IrMn/Fe$_{95}$Gd$_5$/Hf/Co$_{40}$Fe$_{40}$B$_{20}$/Hf multilayer with the exchange bias parallel to the current and the magnetic field applied in plane and perpendicular to the current. (b,c) The second-harmonic Hall signals measured under the same conditions for (b) the IrMn/Hf/Co$_{40}$Fe$_{40}$B$_{20}$/Hf control sample and (c) the IrMn/Fe$_{95}$Gd$_5$/Hf control sample. Field-independent backgrounds have been subtracted from each data set.

To rule out potential experimental artifacts, we prepared two control samples: (i) IrMn(10 nm)/Hf(2 nm)/ $Co_{40}Fe_{40}B_{20}$ (2 nm)/Hf(3 nm) and (ii) IrMn(10 nm)/FeGd(4 nm)/Hf(3 nm), the first having no $Fe_{95}Gd_5$ layer, and the second having no CoFeB layer. We performed the same second-harmonic Hall measurement on each of these samples, with the applied magnetic field perpendicular to the current direction. The sample with no FeGd layer exhibits no field-dependent signal (Fig. 2b). This is as expected, because (as noted above) the orientation of the CoFeB moment transverse to the current should prevent any signals due to spin Hall or Oersted torques and also any thermal signals due to the Nernst effect. Both of these mechanisms should depend only on the behavior of the CoFeB layer; therefore, this measurement allows us to confirm that these signals are indeed absent in our geometry.

The control sample containing only the exchange-biased FeGd layer shows a signal consistent with a dominant contribution from an out-of-plane Oersted torque acting on the FeGd layer and contributing to the Hall voltage via a planar Hall effect, with a small angular misalignment of the exchange bias direction (Fig. 2c) (this signal is non-zero because the exchange-biased FeGd moment is not oriented transverse to the current). The control sample also shows a small signal due to the anomalous Nernst effect (see Supplementary Information). The signal consists of a dip with maximum amplitude centered near *H* = 0 where the FeGd moment is parallel to the current, so that the Oersted torque on the FeGd moment is maximal. The result is therefore qualitatively different than the divergences with a sign change in Fig. 2a. The Oersted torque should be substantially smaller in our full stack than in the FeGd control sample, because in the full stack the part of the Oersted field acting on the FeGd layer that is generated in the Hf and CoFeB layers partially cancels the part of the Oersted field generated in the IrMn layer. We can estimate the size of this signal in our full stack (see Supplementary

Information) from the control measurements, and include this signal as a background in our analysis as described below.

Based on the data in Fig. 2a and from the two control samples (Fig. 2b,c) we conclude that the out-of-plane field-like torque signal observed near $H = 0$ in Fig. 2a is due to a spin current arising from the FeGd layer and acting on the CoFeB layer. This cannot be a conventional spin-orbit torque due to the spin Hall effect, because the conventional spin transfer torque is zero for a magnetic sensor layer oriented perpendicular to the current flow.

To analyze more quantitatively the spin-transfer torque exerted by the FeGd layer acting on the CoFeB layer, we compare to the theory of Taniguchi et al. [17], in which spin-orbit coupling within the ferromagnetic source layer (FeGd) generates a transverse spin current with polarization

$$\vec{\sigma} \propto \left(\hat{y} \cdot \hat{m}_{FeGd}\right)\hat{m}_{FeGd}, \quad (1)$$

where $\hat{m}_{FeGd}$ is a unit vector along the FeGd magnetization direction. Intuitively, we can think of this as a projection of the ordinary SHE spin current onto the magnetization direction. Our system uses an in-plane ferromagnetic sensor layer, and as such is more sensitive to out-of-plane torques than in-plane torques. If this spin polarization interacts with the sensor layer through an effective field $\propto \vec{\sigma}$ (producing an out-of-plane field-like torque), then the effective field produced is $\vec{H}_{FL} = \hat{m}_{FeGd} H_{FL}^0 \sin(\varphi_{FeGd})$. The expected second harmonic signal ($V_H^{2f}$) for the case that the external field $H$ is swept perpendicular to the current-flow direction is (see Supplementary Information)

$$V_H^{2f} = -IR_{PHE}\cos\left(2\varphi_{CoFeB}\right)\frac{H_{FL}^0 \sin\left(\varphi_{FeGd}\right)\sin\left(\varphi_{CoFeB} - \varphi_{FeGd}\right)}{2|H|}. \quad (2)$$

Here $I$ is the applied current, $R_{PHE}$ is planar Hall coefficient of the multilayer due to the CoFeB, and $\varphi_{CoFeB}$ is the angle between the current and the magnetic field, which is either 90° or -90°, following the sign of the applied magnetic field. The $H$ dependence of $V_H^{2f}$ comes from the $H$ dependence of $\varphi_{FeGd}$ and the susceptibility term $1/|H|$. At low field values, it is important to take into account that any angular misalignment of the sample during the annealing will produce a small non-zero value for the orientation of the FeGd magnetization ($\varphi_{FeGd}^0$), so that $\varphi_{FeGd} \approx \varphi_{FeGd}^0 + \tan^{-1}(H/H_{ex})$. Our fits suggest a misalignment $\varphi_{FeGd}^0 \approx 2.7°$.

In performing our fits, we can also account for a small contribution due to the Oersted torque on the FeGd layer, detected in the Hall voltage via the FeGd planar Hall effect. We estimate the size of this background signal generated by the Oersted torque independently (see Supplementary Information). Furthermore, we account for a Nernst signal generated in the CoFeB due to angular misalignment of the sample with respect to the applied field. All of these effects are small; a full discussion (and our procedures for accounting for them) is included in the Supplementary Information.

Figure 3 shows the same second harmonic Hall data as in Fig. 2a with a fit to the theory of Taniguchi et al. [17] (blue line) (see Supplementary Information, fit procedures and parameters). The fit conforms well to the measured data. To illustrate the necessity of taking into account the variation of the transverse spin current $\vec{\sigma}$ on the orientation of the FeGd moment (Eq. 1), we have also performed the fit assuming that $\vec{\sigma}$ is a constant, independent of $\varphi_{FeGd}$ (dashed green line). (This is somewhat artificial, since for the experimental geometry of Fig. 2a one should have $\vec{\sigma}$ = 0 for conventional torques, as explained above.) Any model assuming $\vec{\sigma}$ = constant is qualitatively inconsistent with the measurements, while taking into account the expected variation of $\vec{\sigma}$ with $\varphi_{FeGd}$ accounts well for the nonmonotonic dependence of the signal at positive fields.

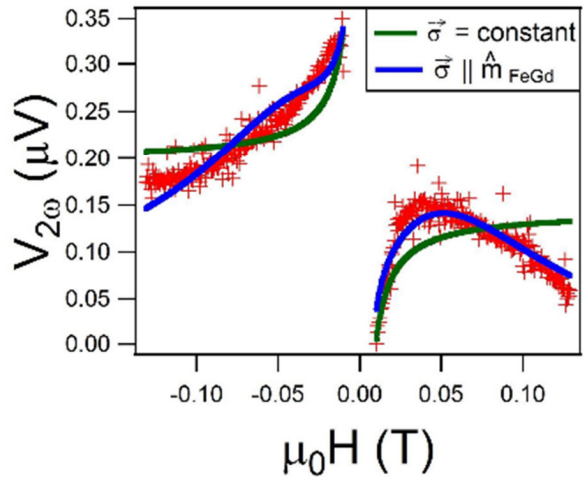

Fig. 3. The same data as in Fig. 2(a), with comparison to the model of Taniguchi et al. [17] (blue line), which assumes that $\vec{\sigma} \propto (\hat{y} \cdot \hat{m}_{FeGd})\hat{m}_{FeGd}$, as well as a fit to a model which assumes that $\vec{\sigma}$ is constant (green dashed line).

We can characterize the strength of the out-of-plane field-like torque generated by the FeGd and acting on the CoFeB in terms of a spin-torque efficiency, $\xi_{FL,AHE}$, such that the spin-current-induced effective field acting on the CoFeB is $H_{FL}^0 = \xi_{FL,AHE} \frac{\hbar}{2e\mu_0 M_s t_{FM}} J_e \sin(\varphi_{FeGd})$, where $J_e$ is the applied charge current density in the FeGd, $\mu_0 M_s$ = 0.90 T is the saturation magnetization of the CoFeB layer based on VSM measurements of a IrMn/Hf/CoFeB/Hf control sample, and $t_{FM}$ is the thickness of the ferromagnetic layer. Fitting the measured signal yields an estimated $\xi_{FL,AHE}$ of -1.8 ± 0.4%. This figure is a lower bound for the magnitude of the field-like spin-torque efficiency that can be generated by the FeGd because we do not account for less-than-perfect interface transparency or the loss of spin current upon transmission through the hafnium spacer [26]. An in-plane torque component may also be present in our samples, but the experimental geometry does not allow an accurate quantitative measurement. The signature of an in-plane torque in a second-harmonic Hall measurement is much less pronounced than the $1/|H|$ divergence for an out-of-plane torque, and is further obscured when the magnitude of the torque is field-dependent. Our best estimate, based on multi-parameter fits, is that the analogous anti-damping torque efficiency is $|\xi_{AD,AHE}|$ < 1.0% (see Supplementary Information).

We note that Humphries et al. [27] have recently pointed out an alternative mechanism whereby an out-of-plane spin-orbit torque might be generated in a ferromagnet/spacer/ferromagnetic multilayer – a spin current generated by spin-orbit interactions with an in-plane spin polarization might precess in the exchange field of the fixed magnetic layer so that when the resulting spin current interacts with the sensor magnetic layer it can apply an out-of-plane anti-damping torque. We can tell that this mechanism is not dominant in our measurement because the out-of-plane torque we measure is a field-like torque, not an anti-damping torque, based on the sign change we observe in the component of the second-harmonic Hall signal proportional to $1/|H|$ upon reversal of the CoFeB magnetization near zero field.

In summary, we have observed experimentally an out-of-plane field-like spin-orbit torque that varies in strength as a function of changes in the direction of the magnetic moment in the source layer. These changes are quantitatively consistent with predictions [16,17] for the variations in the spin current resulting from the anomalous Hall effect within the magnetic layer. The most direct evidence for spin torque from this new mechanism is that it can deflect sensor-layer magnetic moments oriented perpendicular to the charge current flow, whereas both conventional spin Hall torques and the current-induced torque from the Oersted field are identically zero for this geometry. The results we report have been obtained using magnetic layers with in-plane magnetic anisotropy, in order to obtain the best control over magnetic orientations and the fewest competing experimental artifacts, which has the consequence that these measurements are primarily sensitive to an out-of-plane field-like component of spin-orbit torque. We have not yet probed the regime of our primary interest for practical applications – in which $\hat{m}_{\text{source}}$ is tilted out of the sample plane so that the anomalous Hall effect mechanism might apply an anti-damping torque to an active magnetic layer with perpendicular magnetic anisotropy. However, the good agreement between theory [17] and our results with in-plane magnetized layers provides optimism for the pursuit of this goal.

Acknowledgements. We thank G. Stiehl for comments on the manuscript. This work was supported by the National Science Foundation (DMR-1406333) and Western Digital. We made use of the Cornell Nanoscale Facility, a member of the National Nanotechnology Coordinated Infrastructure (NNCI), which is supported by the NSF (ECCS-1542081), and also the Cornell Center for Materials Research Shared Facilities, which are supported by the NSF MRSEC program (DMR-1120296).

Supplementary Information

Reorientable Spin Direction for Spin Current Produced by the Anomalous Hall Effect

Jonathan D. Gibbons,[1] David MacNeill,[1] Robert A. Buhrman,[1] and Daniel C. Ralph[1,2]

[1] Cornell University, Ithaca, NY 14853, USA

[2] Kavli Institute at Cornell for Nanoscale Science, Ithaca, New York 14853, USA

**Derivation of the Second Harmonic Hall Signal for Spin-Orbit Torque Originating from the Anomalous Hall Effect**

We use the Landau-Lifshitz-Gilbert-Slonczewski (LLGS) equation in the macrospin approximation to determine the magnetization orientation of our CoFeB sensor layer ($\vec{m}$), adapting the calculation of Hayashi et al. [M. Hayashi, J. Kim, M. Yamanouchi, and Hideo Ohno, *Quantitative characterization of the S-O torque using harmonic Hall voltage measurements*, Phys. Rev. B **89**, 144425 (2014).]. In the presence of arbitrary spin-current-induced field-like and anti-damping torques, the time-dependent LLGS equation may be written

$$\frac{d\vec{m}}{dt} = -\gamma \vec{m} \times \vec{H}_{tot} + \alpha \vec{m} \times \frac{d\vec{m}}{dt} + \gamma H_{FL} \vec{m} \times \hat{\sigma} + \gamma H_{AD} \vec{m} \times \left( \hat{\sigma} \times \vec{m} \right). \tag{S1}$$

$H_{FL}$ and $H_{AD}$ characterize the strength of the field-like and anti-damping spin-orbit torques. We will assume that for spin-orbit torques generated by the anomalous Hall effect the orientation of the spin current $\hat{\sigma}$ is parallel to the source-layer magnetization $\hat{m}_{FeGd}$, that $H_{FL} = H_{FL}^0 \sin(\varphi_{FeGd})$ and $H_{AD} = H_{AD}^0 \sin(\varphi_{FeGd})$, that the sensor layer magnetization is in-plane, and that the external magnetic field $\vec{H}$ is also applied in-plane. For low-frequency second harmonic Hall measurements, a quasi-steady state condition applies, meaning that the time derivative terms in Eq. (S1) can be taken to be zero. The CoFeB magnetization then follows the total effective field, and since the CoFeB layer is approximately isotropic within the sample plane, $\vec{m}$ should be parallel to $\vec{H} - m_z H_k \hat{z} - H_{FL} \hat{\sigma} - H_{AD} \hat{\sigma} \times \hat{m}$, where $H_k$ is the magnetic anisotropy field of the thin-film CoFeB layer. As a result, the effect of the field-like spin-orbit torque is to rotate the sensor-layer magnetization within the sample plane by an angle

$$\Delta \varphi_{CoFeB} = \frac{H_{FL}^0 \sin(\varphi_{FeGd}) \sin(\varphi_{CoFeB} - \varphi_{FeGd})}{|H|} \tag{S2}$$

relative to its value in the presence of the applied magnetic field but with no spin-orbit torque. This expression assumes that $\left| H_{FL}^0 \right| \ll |H|$. Similarly, the anti-damping torque arising from the anomalous Hall effect should deflect the CoFeB magnetization perpendicular to the sample plane,

$$\Delta\theta_{CoFeB} = \frac{H_{AD}^0 \sin(\varphi_{FeGd})\sin(\varphi_{CoFeB} - \varphi_{FeGd})}{|H| + H_k}. \tag{S3}$$

These deflections will alter the Hall resistance, $R_{XY}$, which can have contributions from the anomalous Hall effect and the planar Hall effect:

$$R_{XY} = \frac{1}{2}R_{AHE}\cos(\theta_{CoFeB}) + \frac{1}{2}R_{PHE}\sin^2(\theta_{CoFeB})\sin(2\varphi_{CoFeB}). \tag{S4}$$

Assuming a sensor layer with in-plane anisotropy (so that in the absence of spin-orbit torques $\theta_{CoFeB} = \pi/2$), the change in the Hall resistance due to the spin-orbit torques is

$$\Delta R_{XY} = R_{PHE}\cos(2\varphi_{CoFeB})\Delta\varphi_{CoFeB} - \frac{1}{2}R_{AHE}\Delta\theta_{CoFeB}. \tag{S5}$$

After mixing with the oscillating applied current $I\sin\omega t$, the oscillating planar Hall resistance will produce a second harmonic signal

$$V_H^{2f} = -\frac{1}{2}IR_{PHE}\cos(2\varphi_{CoFeB})\frac{H_{FL}^0 \sin(\varphi_{FeGd})\sin(\varphi_{CoFeB} - \varphi_{FeGd})}{|H|} + \frac{1}{4}IR_{AHE}\frac{H_{AD}^0 \sin(\varphi_{FeGd})\sin(\varphi_{CoFeB} - \varphi_{FeGd})}{|H| + H_k} \tag{S6}$$

The first term on the right reduces to Eq. (2) in the main text when $\vec{H}$ is applied perpendicular to the current direction and the CoFeB magnetization is saturated parallel to this field, so that $\varphi_{CoFeB} = \pm\pi/2$.

The second term in the right in Eq. (S6), associated with the anti-damping spin-orbit torque, appears to be too small to contribute significantly to our measurement. We have performed our fitting procedure with various values for the size of this anti-damping term, to determine the point at which the fitting becomes poor. This method gives a conservative upper bound of $|\mu_0 H_{AD}^0| \leq 46~\mu T$ for I = 5.2 mA, (corresponding to an anti-damping spin torque efficiency ≤ 1.0 ± 0.3%). However, this does not necessarily reflect an upper bound on the maximum spin torque that can be produced by the FeGd layer, since this bound does not take into account attenuation of the spin current upon transmission through the Hf spacer layer or imperfect spin transmission at interfaces.

**First Harmonic Hall Fits**

At the same time as we measure the second-harmonic Hall voltage, we also measure a first-harmonic Hall voltage. This voltage can be modeled by the sum of planar and anomalous Hall effects.

$$V_H^{1f} = \frac{1}{2} IR_{AHE} \cos(\theta) + \frac{1}{2} IR_{PHE} \sin^2(\theta) \sin(2\varphi)$$

For our samples, we have two different magnetic layers, and the first-harmonic Hall signal will include contributions from both layers:

$$V_H^{1f} = \frac{1}{2} IR_{AHE,CoFeB} \cos(\theta_{CoFeB}) + \frac{1}{2} IR_{PHE,CoFeB} \sin^2(\theta_{CoFeB}) \sin(2\varphi_{CoFeB})$$

$$+ \frac{1}{2} IR_{AHE,FeGd} \cos(\theta_{FeGd}) + \frac{1}{2} IR_{PHE,FeGd} \sin^2(\theta_{FeGd}) \sin(2\varphi_{FeGd}) \ .$$

Here, $R_{PHE,CoFeB}$ and $R_{PHE,FeGd}$ refer to the contributions to the planar Hall resistance due to the CoFeB and FeGd layers, respectively. $R_{AHE,CoFeB}$ and $R_{AHE,FeGd}$ are likewise the contributions to the anomalous Hall resistance due to the CoFeB and FeGd layers. In other sections of this work, $R_{PHE}$ and $R_{AHE}$ are used to refer to the planar Hall resistance and anomalous Hall resistance, respectively, of the relevant sensor material that is being discussed. For the purpose of quantitative analysis, we need to distinguish the two, and know separately the size of the two $R_{PHE}$ terms in particular.

The first-harmonic data taken concurrently with the primary second-harmonic measurements can be fit to the following equation:

$$V_H^{1f} = C_{AHE} H + C_{PHE} \sin\left(2 \arctan\left[\frac{H}{H_{EB}}\right] + 2\varphi_{FeGd}^0\right) + C_{offset}$$

In the above equation, the fit parameters are as follows:

$C_{PHE} = \frac{1}{2} IR_{PHE,FeGd}$ is the size of the planar Hall effect in FeGd.

$C_{AHE}$ is a constant that depends on the out-of-plane component of the magnetic field. The magnetic field is predominantly in-plane, but does have an out-of-plane component that contributes a linear signal to the first-harmonic Hall voltage.

$C_{offset}$ is a constant an offset voltage.

$H_{EB}$ is the exchange bias field amplitude.

$\varphi_{FeGd}^0$ is a misalignment angle. This fit is very insensitive to this angle, and does not give an accurate estimate of its value.

The fit to the first-harmonic Hall data therefore gives us the value of the exchange bias field, as well as the size of the planar Hall signal in the FeGd layer.

To determine the size of the planar Hall effect in the CoFeB layer, we perform a first-harmonic planar Hall measurement on the IrMn/Hf/CoFeB control sample. To apply the result to the full IrMn/FeGd/Hf/CoFeB multilayer, we take care to account for current shunting by correcting the applied current so that the same current flows in the CoFeB layer of the control sample as in the full multilayer.

**Signal due to the Oersted Field acting on the FeGd layer**.

For our geometry, the applied magnetic field is parallel to the current-induced Oersted field (both are perpendicular to the current) and thus the CoFeB magnetization is also parallel to the Oersted field. Therefore the Oersted field creates no rotation of the CoFeB layer. However, due to the exchange bias from the IrMn layer the FeGd magnetization is not in general parallel to the Oersted field, and thus the FeGd layer may exhibit a planar Hall second-harmonic signal due to the Oersted-generated rotation of the FeGd magnetization (e.g. Fig. 2c). The Oersted field competes against both the applied magnetic field, $H$, and the exchange bias field, $H_{EB}$, which are perpendicular to one another. Hence, the Oersted-field-induced rotation in a macrospin approximation is

$$\Delta\varphi_{FeGd} = \frac{H_{Oe}\cos(\varphi_{FeGd})}{\sqrt{H^2 + H_{EB}^2}}. \tag{S7}$$

This leads to a second-harmonic Hall signal

$$V_H^{2f} = IR_{PHE}\cos(2\varphi_{FeGd})\frac{H_{Oe}\cos(\varphi_{FeGd})}{2\sqrt{H^2 + H_{EB}^2}}. \tag{S8}$$

We can probe the size of this signal in our primary IrMn(10 nm)/FeGd(4 nm)/Hf(2 nm)/CoFeB(2 nm)/Hf(3 nm) multilayer by making comparison to the IrMn(10 nm)/FeGd(4 nm)/Hf(3 nm) control sample, for which this mechanism makes the primary contribution to the second-harmonic Hall signal (see Fig S1a). However, in the control sample the Oersted field is generated almost exclusively by current flow in the IrMn layer, while in the primary sample there are two additional current-carrying layers (the Hf spacer and the CoFeB layer) that will produce an Oersted field that will partially cancel the Oersted field from the IrMn layer in that sample. The Hall signal arising from the Oersted-induced reorientation of the FeGd layer will therefore be smaller in the primary sample than in the control.

In order to estimate this signal, we have calculated the Oersted field in the primary sample based on the resistivity measurements of the individual layers. The first harmonic signal from the primary experiment is due to the planar Hall effect in FeGd. Therefore, our fit to the data also gives us the component of the planar Hall resistance that is due to the FeGd. By combining this value with the calculated Oersted field, we can predict the size of the Oersted parasitic signal. The result was treated as a smooth fixed background in our fits to the second harmonic Hall signal as a function of applied field. Figure S2 shows the results of the fits with and without this estimated background signal included for our primary IrMn(10 nm)/FeGd(4 nm)/Hf(2 nm)/CoFeB(2 nm)/Hf(3 nm) multilayers. The background is small enough (and smooth enough) that it does not influence any of the conclusions of our paper.

This background correction is of course only approximately correct because the resistivities of the materials in the multilayer will not necessarily be exactly the same as in separate individual layers. In order to check this, we calculate the expected Oersted parasitic signal for our IrMn(10 nm)/FeGd(4 nm)/Hf(2 nm) control sample, and confirm that the size is consistent with what is observed in the control sample at 30 K.

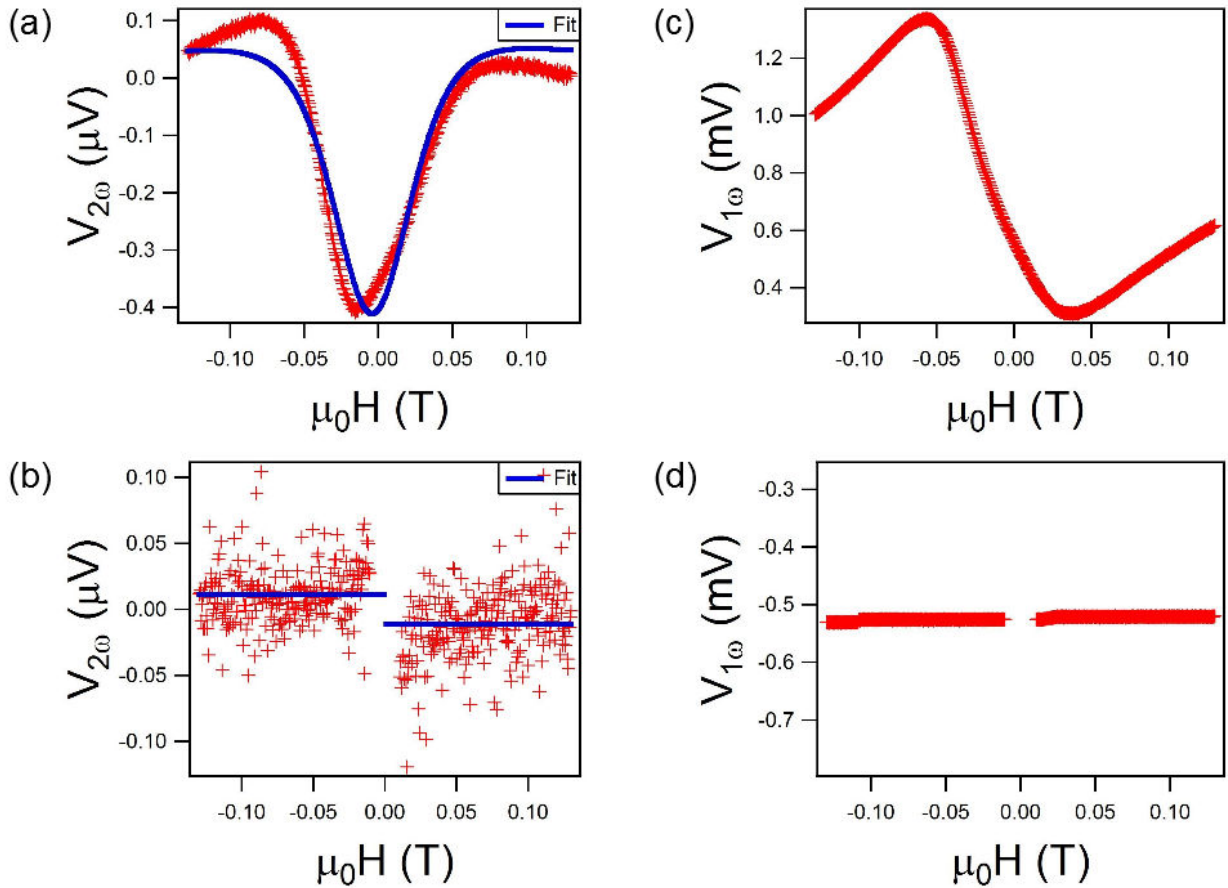

Fig. S1 (a) Second harmonic Hall data for the IrMn/FeGd/Hf control sample, with fit using the calculated value for the Oersted field. (b) Second harmonic Hall data for the IrMn/Hf/CoFeB/Hf control sample, with fit to a misalignment Nernst term. (c) First harmonic Hall data for the IrMn/FeGd/Hf control sample. (d) First harmonic Hall data for the IrMn/Hf/CoFeB/Hf control sample.

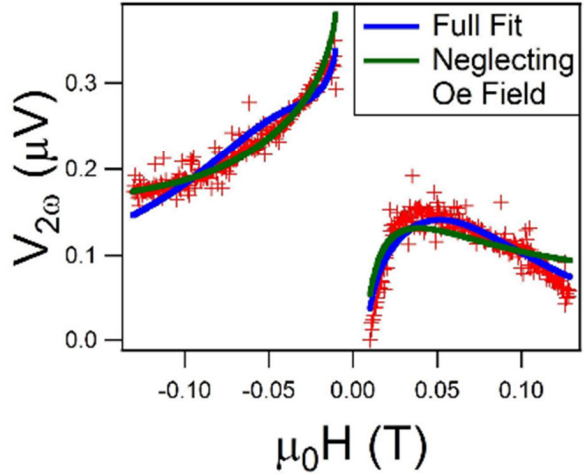

Fig S2. A comparison of our full fit to a fit done neglecting a small, smooth background due to the Oersted field acting on the FeGd layer and producing a second harmonic signal via the planar Hall effect.

**Signals due to the Anomalous Nernst Effect**

Two other sources of parasitic signals may exist in our measurements – first, a potential anomalous Nernst signal generated by a vertical thermal gradient in the CoFeB layer. Such a signal only occurs due to a small misalignment of the applied magnetic field in our experimental setup. This signal generates a very small constant term that changes sign when the CoFeB layer switches. Such a misalignment may also cause an Oersted-field generated planar Hall voltage in the CoFeB. However, this signal will fall off quickly with applied field and will therefore make no significant contribution to our fits. We observe a small ANE signal in the CoFeB control sample due to misalignment of the applied field; we include this term in our primary fit, but it is seen to be fairly small (see fitting parameters).

Second, we may expect to see an anomalous Nernst signal generated in the FeGd layer, which varies smoothly as $V_{ANE} \propto \cos(\varphi_{FeGd})$. A fit of the second harmonic data measured in the IrMn/FeGd control sample sets a bound of < 50 nV at 30 K for the $V_{AHE}$ contribution. The size of the anomalous Nernst signal in the FeGd is small enough that it makes no meaningful contribution to the control samples, and so we do not include it in the fit to our primary results.

**Fit procedures and parameters.**

We determined the following fixed parameters by independent measurements:

*In the CoFeB layer*:

$IR_{PHE} = 0.17 \pm 0.02$ mV was determined by measuring the Hall voltage using a lock-in amplifier while rotating the direction of an in-plane magnetic field applied to the CoFeB control sample, maintaining the same current through the CoFeB layer as in the primary sample.

*In the FeGd layer*:

$IR_{PHE}\mu_0 H_{Oe} = -23 \pm 2$ nV T was calculated, and separately confirmed using the FeGd control, as outlined above.

$\mu_0 H_{ex}$ = 0.071 ± 0.001 T, measured with a linear field first harmonic measurement with the field perpendicular to the exchange bias.

After accounting for the signal due to the Oersted field acting on the FeGd layer as discussed above, we fit the measured data to Eq. (S6) (with $H_{AD}^0 = 0$) using three adjustable parameters plus an overall offset voltage. The fit equation is as follows.

$$V^{2f} = -IR_{PHE}\frac{H_{FL}^0 \sin(2\varphi_{FeGd})}{4H} + V_{Offset} + V_{ANE,Misalignment}\,\text{sgn}(H) + IR_{PHE}\cos(2\varphi_{FeGd})\frac{H_{Oe}\cos(\varphi_{FeGd})}{2\sqrt{H^2 + H_{EB}^2}},$$

with $\varphi_{FeGd} \approx \varphi_{FeGd}^0 + \tan^{-1}(H/H_{ex})$.

Using this fit, we determine the following values.
$H_{FL}^0$ = 0.28 ± 0.01 mT for *I* = 5.2 mA.

$\varphi_{FeGd}^0$ = 2.7° ± 0.2°

$V_{ANE,Misalignment}$ = 40 ± 2 nV.